\pdfoutput=1
\documentclass{quantumview}
\usepackage[utf8]{inputenc}
\usepackage[english]{babel}
\usepackage[T1]{fontenc}
\usepackage{amsmath}
\usepackage{hyperref}
\usepackage{fixltx2e}
\usepackage[numbers,sort&compress]{natbib}
\usepackage{tabularx}

\begin{document}

\title{Robust multi-party semi-quantum private comparison protocols with decoherence-free states against collective noise}

\author{Lihua Gong}\email{lhgong@ncu.edu.cn}
\author{Zhenyong Chen}
\affiliation{Department of Electronic Information Engineering, Nanchang University, Nanchang 330031, China}
\author{Liguo Qin}
\author{Jiehui Huang}
\affiliation{School of Mathematics, Physics and Statistics, Shanghai University of Engineering Science, Shanghai 201620, China}

\maketitle

\begin{abstract}
\noindent Based on decoherence-free states, two multi-party semi-quantum private comparison protocols are proposed to counteract collective noises. One could resist the collective-dephasing noise well, whereas the other could resist the collective-rotation noise. Multiple classical participants could compare their secret information by performing the proposed protocols once. It is manifested that the proposed protocols could resist both external attacks and internal attacks. Besides, the operations of our protocols were simulated on the IBM Quantum Experience.
\end{abstract}

\section{Introduction}
As an intriguing application of quantum secure multiparty computing \cite{1}, quantum private comparison (QPC) which could compare participants’ private information while keeping their secrets has gained wide attention in recent years. The first QPC protocol based on decoy particles was invented by Yang et al \cite{2}. Afterward, QPC protocols combining various quantum technologies including entanglement-swapping \cite{3,4}, quantum walk \cite{5}, and bit-flipping \cite{6} have been investigated. 

Multi-party QPC (MQPC) has made tremendous progress in that it could compare multiple participants’ privacies within one execution of protocols. The MQPC protocol was presented based on multi-particle GHZ class states \cite{7}. Subsequently, two MQPC protocols were put forward with multi-level entangled states in distributed model and in traveling model, respectively \cite{8}. Some MQPC protocols were also designed with entanglement swapping \cite{9,10}. An MQPC without authentication channels and pre-shared keys between participants was proposed with the help of two third parties \cite{11}. By considering multi-dimensional single particles as information carriers, two MQPC protocols were expatiated to achieve the size comparison among multiple participants’ secrets \cite{12}. Laterly, an MQPC protocol based on qubit shift operation was discussed \cite{13}. A class of effective MQPC protocols was presented to flexibly select quantum states \cite{14}. Recently, a participant-independent MQPC protocol was dwelt on with  $d$-dimensional Bell states \cite{15}.
It should be noted that all participants in the above-mentioned quantum private comparison protocols should possess sufficient quantum abilities. However, the expensive quantum devices may be not always available to the participants. After the conception of semi-quantum was initiated \cite{16}, a set of semi-quantum private comparison (SQPC) protocols have been presented \cite{17,18,19}. Recently, the first multi-party SQPC (MSQPC) protocol was designed with Bell sates \cite{20}.  

However, the MSQPC scheme in \cite{20} could only work under the ideal noiseless environments. In fact, the negative influences of noises couldn’t be neglected since the particles may be disturbed by noises during transmission. Various noise-resisting QPC protocols were researched \cite{21,22,23}. Usually, the noises associated with quantum channels could be considered as collective noises. To overcome the collective noises, one of the most effective approaches is to encode and transmit secrets with the decoherence-free (DF) states. By adopting the DF states as information carriers, two robust MSQPC protocols will be designed to effectively prevent collective-dephasing noise and collective-rotation noise, respectively.
The remainder of this paper is organized as follows. In Section \ref{sec.2}, two robust MSQPC protocols against collective noises are described in detail. In Section \ref{sec.3}, the robustness of the two protocols in resisting external attacks and internal ones are presented. In Section \ref{sec.4}, the operations of the two protocols are implemented with the IBM quantum computer. In Section \ref{sec.5}, the two presented MSQPC protocols are evaluated with some baseline protocols. Ultimately, a succinct conclusion is shown in Section \ref{sec.6}.

\section{Two robust MSQPC protocols against collective noises}\label{sec.2}
The collective-dephasing noise ${U_{dp}}$ can be modeled as
\begin{equation}
{U_{dp}}\left| 0 \right\rangle  = \left| 0 \right\rangle, {U_{dp}}\left| 1 \right\rangle  = {e^{{\rm{i}}\theta }}\left| 0 \right\rangle.
\end{equation}
where $\theta $ denotes the noise parameter fluctuating with time \cite{23}. The logical qubits  $\left| {0{}_{dp}} \right\rangle$ and $\left| {1{}_{dp}} \right\rangle$ can be expressed by the DF states $\left| {01} \right\rangle $ and $\left| {10} \right\rangle $,
\begin{equation}
\left| {0{}_{dp}} \right\rangle  = \left| {01} \right\rangle, \left| {1{}_{dp}} \right\rangle  = \left| {10} \right\rangle.
\end{equation}

It is easily found that the two logical qubits could resist the collective-dephasing noise, since they have the same phase shift, i.e., ${\rm{i}}\theta$,
\begin{equation}
{U_{dp}}\left| {0{}_{dp}} \right\rangle  = {e^{{\rm{i\theta }}}}\left| {01} \right\rangle, {U_{dp}}\left| {1{}_{dp}} \right\rangle  = {e^{{\rm{i\theta }}}}\left| {10} \right\rangle.
\end{equation}
The superposition states $\left| { + {}_{dp}} \right\rangle  = \frac{1}{{\sqrt 2 }}\left( {\left| {0{}_{dp}} \right\rangle  + \left| {1{}_{dp}} \right\rangle } \right) = \frac{1}{{\sqrt 2 }}\left( {\left| {01} \right\rangle  + \left| {10} \right\rangle } \right) = \left| {{\psi ^ + }} \right\rangle$ and $\left| { - {}_{dp}} \right\rangle  = \frac{1}{{\sqrt 2 }}\left( {\left| {0{}_{dp}} \right\rangle  - \left| {1{}_{dp}} \right\rangle } \right) = \frac{1}{{\sqrt 2 }}\left( {\left| {01} \right\rangle  - \left| {10} \right\rangle } \right) = \left| {{\psi ^ - }} \right\rangle$ are also invariant towards the collective-dephasing noise. Besides, ${Z_{dp}}:\left\{ {\left| {0{}_{dp}} \right\rangle ,\left| {1{}_{dp}} \right\rangle } \right\}$ and ${X_{dp}}:\left\{ {\left| { + {}_{dp}} \right\rangle ,\left| { - {}_{dp}} \right\rangle } \right\}$ are two measurement bases under the collective-dephasing noise.

The collective-rotation noise ${U_r}$ can be represented as
\begin{equation}
{U_r}\left| 0 \right\rangle  = \cos \theta \left| 0 \right\rangle  + \sin \theta \left| 0 \right\rangle, {U_r}\left| 1 \right\rangle  =  - \sin \theta \left| 0 \right\rangle  + \cos \theta \left| 1 \right\rangle.
\end{equation}
The logical qubits $\left| {0{}_r} \right\rangle$ and $\left| {1{}_r} \right\rangle$ are defined by the DF states  $\frac{1}{{\sqrt 2 }}\left( {\left| {00} \right\rangle  + \left| {11} \right\rangle } \right) = \left| {{\phi ^ + }} \right\rangle $ and $\frac{1}{{\sqrt 2 }}\left( {\left| {01} \right\rangle  - \left| {10} \right\rangle } \right) = \left| {{\psi ^ - }} \right\rangle $,
\begin{equation}
\left| {0{}_r} \right\rangle  = \frac{1}{{\sqrt 2 }}\left( {\left| {00} \right\rangle  + \left| {11} \right\rangle } \right),\left| {1{}_r} \right\rangle  = \frac{1}{{\sqrt 2 }}\left( {\left| {01} \right\rangle  - \left| {10} \right\rangle } \right).
\end{equation}

The two logical qubits can be immune to the collective-rotation noises,
\begin{equation}
{U_r}\left| {0{}_r} \right\rangle  = \left| {0{}_r} \right\rangle, {U_r}\left| {1{}_r} \right\rangle  = \left| {1{}_r} \right\rangle. 
\end{equation}
The superposition states $\left| { + {}_r} \right\rangle  = \frac{1}{{\sqrt 2 }}\left( {\left| {0{}_r} \right\rangle  + \left| {1{}_r} \right\rangle } \right) = \frac{1}{{\sqrt 2 }}\left( {\left| {{\phi ^ + }} \right\rangle  + \left| {{\psi ^ - }} \right\rangle } \right)$ and $\left| { - {}_r} \right\rangle  = \frac{1}{{\sqrt 2 }}\left( {\left| {0{}_r} \right\rangle  - \left| {1{}_r} \right\rangle } \right) = \frac{1}{{\sqrt 2 }}\left( {\left| {{\phi ^ + }} \right\rangle  - \left| {{\psi ^ - }} \right\rangle } \right)$ could also resist this kind of noises.  Moreover, ${Z_r}:\left\{ {\left| {0{}_r} \right\rangle ,\left| {1{}_r} \right\rangle } \right\}$ and ${X_r}:\left\{ {\left| { + {}_r} \right\rangle ,\left| { - {}_r} \right\rangle } \right\}$ are two mutually nonorthogonal bases under the collective-rotation noise. 

With the aid of a semi-honest party, i.e., TP to control qubits, the privacies of $n$ classical participants ${P_1},{P_2}, \cdot  \cdot  \cdot ,{P_n}$ can be compared securely, where  ${P_i}$ $\left( {i = 1,2, \cdot  \cdot  \cdot ,n} \right)$  owns private binary information ${X_i} = ({x_{i,0}},{x_{i,1}}, \cdot  \cdot  \cdot ,{x_{i,l - 1}})$ , ${x_{i,j}} \in \{ 0,1\}$ , $j = 0,1, \cdot  \cdot  \cdot ,l - 1$ . The classical participants in our MSQPC protocols are allowed to execute the operations below.

\noindent(1)	CTRL: Forwarding the particles directly;

\noindent(2)	SIFT: Measuring the particles in $Z \otimes Z$  basis $\left\{ {\left| {00} \right\rangle \left| {01} \right\rangle ,\left| {10} \right\rangle ,\left| {11} \right\rangle } \right\}$ , preparing fresh qubits in the same states and distributing them back;

\noindent(3)	Reordering the particles with different delay lines.

\subsection{MSQPC protocol against the collective-dephasing noises}
Classical participants should share one common key   via a secure SQKD protocol against the collective-dephasing noises beforehand \cite{24}, where $K = ({K_0},{K_1},...,{K_{l - 1}})$ , ${K_j} \in \{ 0,1\} $.

\noindent\textbf{Step 1} TP randomly prepares $n$ sequences ${S_{dp(1)}},{S_{dp(2)}},...,{S_{dp(n)}}$ , each sequence contains $4l(1 + \delta )$ particles in ${Z_{dp}}$ basis and $l(1 + \delta )$ particles in  ${X_{dp}}$ basis, where $\delta $ is a fixed positive fraction. Then TP distributes ${S_{dp(i)}}$ to ${P_i}$.

\noindent\textbf{Step 2} After receiving sequence ${S_{dp(i)}}$, ${P_i}$ executes CTRL or SIFT randomly on the received particles. According to the encoding rule such that $\left| {01} \right\rangle  \to 0$ , $\left| {10} \right\rangle  \to 1$ , ${P_i}$ writes down the classical bits when he executes the SIFT. ${P_i}$ reorders the particles with different delay lines and distributes the new sequence $S_{dp(i)}^{'}$ to TP. 

\noindent\textbf{Step 3} Upon receiving each particle from ${P_i}$, TP announces the particle locations in ${Z_{dp}}$ basis in the original sequence to ${P_i}$. ${P_i}$ publishes the correct orders of sequence $S_{dp(i)}^{'}$ and the operations he performed on each particle pair. TP rearranges the order of sequence $S_{dp(i)}^{'}$. According to the initial bases of particle pairs and the information of ${P_i}$, TP takes the corresponding actions, as shown in Table \ref{Table 1}.

\begin{table}[!ht]
	\begin{center}
	\caption{Actions of TP in the MSQPC protocol against the collective-dephasing noises}
	\begin{tabular}{  m{1.5cm}<{\centering} m{2.6cm}<{\centering}  m{2cm}<{\centering}  m{5cm} } 	
		\hline
		   & Initial basis of particle pair &  Operation of ${P_i}$ &  Action of TP \\ 
		 \hline
	\textbf{Case 1} & ${Z_{dp}}$ $\left( {{X_{dp}}} \right)$ & CTRL & \textbf{Action 1}: measuring the particles with ${Z_{dp}}$ $\left( {{X_{dp}}} \right)$  basis \\ 
    \textbf{Case 2} & ${Z_{dp}}$ & SIFT & \textbf{Action 2}: counting the number of the particles \\ 
    \textbf{Case 3} & ${X_{dp}}$ & SIFT & \textbf{Action 3}: dropping out the particles directly \\ 
         \hline	
         \label{Table 1}
\end{tabular}
\end{center}
\end{table}

\noindent\textbf{Case 1} If ${P_i}$ selects to execute CTRL, then TP will take \textbf{Action 1}. TP could calculate the error rate by comparing the measurement results with the initial states. If the error rate exceeds the tolerable limit, then the quantum channel is supposed to be insecure, thus the protocol will be terminated. Otherwise, the protocol will continue to the next step. 

\noindent\textbf{Case 2} If ${P_i}$ executes SIFT and the initial basis of particle pair is ${Z_{dp}}$, then TP will take \textbf{Action 2}. The protocol will be terminated if the number of particle groups is less than $2l$.

\noindent\textbf{Case 3} If ${P_i}$ executes SIFT and the initial basis of particle pair is ${X_{dp}}$, then TP will take \textbf{Action 3}.

\noindent\textbf{Step 4} To verify whether TP is honest or not, ${P_i}$ randomly selects $l$ particle groups in \textbf{Case 2} as test qubits and declares the locations of the test qubits to TP. TP tells the initial states of test qubits to ${P_i}$.  If the initial state of the test qubit is $\left| {{0_{dp}}} \right\rangle $, then the classical bit  ${P_i}$ recoded in \textbf{Step 2} will be 0. Otherwise, the classical bit will be 1. Subsequently, ${P_i}$ computes the error rate. If the error rate is less than the threshold, the protocol will go to the next step. Otherwise, the protocol will be halted.

\noindent\textbf{Step 5} ${P_i}$ selects $l$ particle groups from the existing particles and recodes the classical bits of the corresponding particles as ${m_i}$. Then ${P_i}$ publishes the corresponding locations to TP. ${P_i}$ computes ${r_{i,j}} = {K_j} \oplus {x_{i,j}} \oplus {m_{i,j}}$ and publishes ${r_{i,j}}$ to TP via a classical channel, where $\oplus$ is the exclusive-OR operation. 

\noindent\textbf{Step 6} TP could obtain ${M_{i,j}}$ easily since he knows the initial states and the positions of the selected particles. Then TP calculates and stores ${u_{i,j}} = {M_{i,j}} \oplus {r_{i,j}}$. Once obtaining ${u_{1,j}},{u_{2,j}},...,{u_{n,j}}$, TP could compare the privacies of $n$ classical participants by calculating ${C_j} = ({u_{1,j}} \oplus {u_{2,j}}) + ({u_{2,j}} \oplus {u_{3,j}}) + ... + ({u_{n - 1,j}} \oplus {u_{n,j}})$. Only all elements in $C = \{ {C_0},{C_1},...,{C_{l - 1}}\}$ are 0 can TP concludes that all the classical participants’ secrets are equal.

Apparently, ${M_{i,j}}$ is equal to ${m_{i,j}}$, thus TP could derive ${M_{i,j}} \oplus {m_{i,j}} = 0$ easily. ${u_{1,j}} \oplus {u_{2,j}}$, ${u_{2,j}} \oplus {u_{3,j}}$ ,..., ${u_{n-1,j}} \oplus {u_{n,j}}$ could be simplified as
\begin{equation}
\begin{array}{c}
{u_{1,j}} \oplus {u_{2,j}} = {M_{1,j}} \oplus {K_j} \oplus {x_{1,j}} \oplus {m_{1,j}} \oplus {M_{2,j}} \oplus {K_j} \oplus {x_{2,j}} \oplus {m_{2,j}} \\= {x_{1,j}} \oplus {x_{2,j}};\\
{u_{2,j}} \oplus {u_{3,j}} = {M_{2,j}} \oplus {K_j} \oplus {x_{2,j}} \oplus {m_{2,j}} \oplus {M_{3,j}} \oplus {K_j} \oplus {x_{3,j}} \oplus {m_{3,j}} \\= {x_{2,j}} \oplus {x_{3,j}};\\
...\\
{u_{n - 1,j}} \oplus {u_{n,j}} = {M_{n - 1,j}} \oplus {K_j} \oplus {x_{n - 1,j}} \oplus {m_{n - 1,j}} \oplus {M_{n,j}} \oplus {K_j} \oplus {x_{n,j}} \oplus {m_{n,j}} \\= {x_{n - 1,j}} \oplus {x_{n,j}}.
\end{array}
\end{equation}
Therefore, ${C_j}$ could be obtained as
\begin{equation}
\begin{array}{c}
{C_j} = ({u_{1,j}} \oplus {u_{2,j}}) + ({u_{2,j}} \oplus {u_{3,j}}) + ... + ({u_{n - 1,j}} \oplus {u_{n,j}})\\
= ({x_{1,j}} \oplus {x_{2,j}}) + ({x_{2,j}} \oplus {x_{3,j}}) + ... + ({x_{n - 1,j}} \oplus {x_{n,j}}).
\end{array}
\end{equation}
The output of the MSQPC protocol against the collective-dephasing noises is accurate.

\subsection{MSQPC protocol against the collective-rotation noises}

Similar to the MSQPC protocol against the collective-dephasing noises, all classical participants should share one common key ${K^*}$ via a SQKD protocol against the collective-rotation noises \cite{25}, where ${K^*} = (K_0^*,K_1^*,...,K_{l - 1}^*)$, $K_j^* \in \{ 0,1\} $. 

\noindent\textbf{Step 1*} TP generates $n$ quantum sequences ${S_{r(1)}},{S_{r(2)}},...,{S_{r(n)}}$, which involves $4l(1 + \delta )$ particles in ${Z_r}$ basis and $l(1 + \delta )$ ones in ${X_r}$ basis, respectively. Afterward, TP distributes sequence ${S_{r(i)}}$ to ${P_i}$.

\noindent\textbf{Step 2*} ${P_i}$ executes CTRL or SIFT randomly on the receiving particles. According to the rule such that $\left| {00} \right\rangle /\left| {11} \right\rangle  \to 0$, $\left| {01} \right\rangle /\left| {10} \right\rangle  \to 1$, ${P_i}$ encodes the classical bits when he executes the SIFT operation. Then ${P_i}$ reorders the particles with different delay lines. After ${P_i}$ finished all operations, a new sequence $S_{r(i)}^{'}$ is generated. Hereafter, ${P_i}$ distributes  $S_{r(i)}^{'}$ to TP.

\noindent\textbf{Step 3*} TP stores the particles in sequence $S_{r(i)}^{'}$ and publishes the locations of the particles in ${Z_r}$ basis in the origin sequence to ${P_i}$. ${P_i}$ announces the correct orders of  $S_{r(i)}^{'}$ and the operations he performed on each particle pair. Then TP rearranges the order of the sequence. According to the initial bases of the particle pairs and the operations of ${P_i}$, TP executes the relevant actions, as shown in Table \ref{Table 2}.

\begin{table}[!ht]
	\begin{center}
		\caption{Actions of TP in the MSQPC protocol against the collective-rotation noises}
		\begin{tabular}{  m{1.5cm}<{\centering} m{2.6cm}<{\centering}  m{2cm}<{\centering}  m{5cm} } 	
			\hline
			& Initial basis of particle pair &  Operation of ${P_i}$ &  Action of TP \\ 
			\hline
			\textbf{Case 1*} & ${Z_{r}}$ $\left( {{X_{r}}} \right)$ & CTRL & \textbf{Action 1*}: measuring the particles with ${Z_{r}}$ $\left( {{X_{r}}} \right)$  basis \\ 
			\textbf{Case 2*} & ${Z_{r}}$ & SIFT & \textbf{Action 2*}: counting the number of the particles \\ 
			\textbf{Case 3*} & ${X_{r}}$ & SIFT & \textbf{Action 3*}: dropping out the particles directly \\ 
			\hline	
			\label{Table 2}
		\end{tabular}
	\end{center}
\end{table}

\noindent\textbf{Case 1*} If ${P_i}$  performs CTRL, then TP will execute \noindent\textbf{ Action 1*}. TP computes the error rate by contrasting the measurement outcomes with the initial states. Only the error rate is less than the tolerable limit will the protocol continue.

\noindent\textbf{Case 2*} If ${P_i}$ executes SIFT and the initial basis of particle pair is ${Z_r}$, then TP performs \textbf{Action 2*}. The protocol will continue if the number of particle groups is higher than $2l$.

\noindent\textbf{Case 3*} If ${P_i}$ performs SIFT and the initial basis of particle pair is ${X_r}$, TP will take \textbf{Action 3*}.

\noindent\textbf{Step 4*} ${P_i}$ randomly selects half of particle groups in \textbf{Case 2*} as test bits and announces their positions to TP via a classical channel. TP tells the initial states of test bits to ${P_i}$. If the initial state of test qubit is $\left| {{0_r}} \right\rangle $ $\left( {\left| {{1_r}} \right\rangle } \right)$, then the classical bit   encoded in \textbf{Step 2*} will be 0 (1). Therefore, ${P_i}$ could know whether TP is honest or not. If TP is honest, the protocol will continue. Otherwise, it will be aborted.

\noindent\textbf{Step 5*} ${P_i}$ selects $l$ particle groups from the existing particles and codes the classical bits of the selected particles as $m_i^*$. Then ${P_i}$ calculates $r_{i,j}^* = K_j^* \oplus {x_{i.j}} \oplus m_{i,j}^*$ and publishes the positions to TP. Ultimately, ${P_i}$ tells $r_{i,j}^*$ to TP through a classical authenticated channel. 

\noindent\textbf{Step 6*} Since TP knows the initial states and the locations of the selected particles, he could obtain $M_{i,j}^*$ easily. Then TP calculates and stores $u_i^* = M_i^* \oplus r_i^*$. On obtaining $u_{1,j}^*,u_{2,j}^*,...,u_{2,j}^*$, TP calculates $C_j^* = (u_{1,j}^* \oplus u_{2,j}^*) + (u_{2,j}^* \oplus u_{3,j}^*) + ... + (u_{n - 1,j}^* \oplus u_{n,j}^*)$. If all the elements in ${C^*} = \{ C_0^*,C_1^*,...,C_{l - 1}^*\}$ are 0, then TP will announce the privacies of all classical participants are equal.

Evidently, $M_{i,j}^*$ is identical to $m_{i,j}^*$. Therefore, $M_{i,j}^* \oplus m_{i,j}^* = 0$ could be easily deduced by TP. ${u_{1,j}} \oplus {u_{2,j}}$, ${u_{2,j}} \oplus {u_{3,j}}$,…, ${u_{n-1,j}} \oplus {u_{n,j}}$ could be deduced as

\begin{equation}
\begin{array}{c}
u_{1,j}^* \oplus u_{2,j}^* = M_{1,j}^* \oplus K_j^* \oplus {x_{1.j}} \oplus m_{1,j}^* \oplus M_{2,j}^* \oplus K_j^* \oplus {x_{2.j}} \oplus m_{2,j}^* \\= {x_{1.j}} \oplus {x_{2.j}};\\
u_{2,j}^* \oplus u_{3,j}^* = M_{2,j}^* \oplus K_j^* \oplus {x_{2.j}} \oplus m_{2,j}^* \oplus M_{3,j}^* \oplus K_j^* \oplus {x_{3.j}} \oplus m_{3,j}^* \\= {x_{2.j}} \oplus {x_{3.j}};\\
...\\
u_{n - 1,j}^* \oplus u_{n,j}^* = M_{n - 1,j}^* \oplus K_j^* \oplus {x_{n - 1.j}} \oplus m_{n - 1,j}^* \oplus M_{n,j}^* \oplus K_j^* \oplus {x_{n.j}} \oplus m_{n,j}^* \\= {x_{n - 1.j}} \oplus {x_{n.j}}.
\end{array}
\end{equation}
$C_j^*$ could be generated as

\begin{equation}
\begin{array}{c}
C_j^* = (u_{1,j}^* \oplus u_{2,j}^*) + (u_{2,j}^* \oplus u_{3,j}^*) + ... + (u_{n - 1,j}^* \oplus u_{n,j}^*)\\
= ({x_{1,j}} \oplus {x_{2,j}}) + ({x_{2,j}} \oplus {x_{3,j}}) + ... + ({x_{n - 1,j}} \oplus {x_{n,j}}).
\end{array}
\end{equation}
The output of the MSQPC protocol against the collective-rotation noises is correct.

\section{Security analyses}\label{sec.3}
\subsection{Outside attack}
An outside eavesdropper Eve desires to acquire the secrets of classical participant ${P_i}$ . The possible attack strategies exploited by Eve are mainly the intercept-resend attack, the measure-resend attack, the entanglement attack and the Trojan horse attack.

\subsubsection{Intercept-resend attack}

Eve intercepts sequence ${S_{dp(i)}}$ $\left( {{S_{r(i)}}} \right)$  transmitted from TP to ${P_i}$  and sends a fake sequence ${S_E}$  she generated in advance to ${P_i}$ . Suppose that the particles in ${S_E}$  are in $\left| {{0_{dp}}} \right\rangle $ $\left( {\left| {{0_r}} \right\rangle } \right)$ . Fortunately, TP has a chance to discover Eve’s illegal actions in \textbf {Case 1 (Case 1*)}. Eve couldn’t be perceived if the genuine particle pair sent from TP is also in $\left| {{0_{dp}}} \right\rangle $ $\left( {\left| {{0_r}} \right\rangle } \right)$. If the genuine particle pair sent from TP is in $\left| {{1_{dp}}} \right\rangle $ $\left( {\left| {{1_r}} \right\rangle } \right)$, Eve will be perceived. If the genuine particle pair sent from TP is in ${X_{dp}}$ $\left( {{X_r}} \right)$ basis, the probability of perceiving attack will be $\frac{1}{2}$ . The particles in  $\left| {{1_{dp}}} \right\rangle $ $\left( {\left| {{1_r}} \right\rangle } \right)$ account for $\frac{2}{5}$ of the total particles and the particles in  ${X_{dp}}$ $\left( {{X_r}} \right)$ basis account for $\frac{1}{5}$ of the total particles. Besides, ${P_i}$ will execute CTRL operation with a probability of $\frac{1}{2}$. Hence, when a particle group is attacked, the probability of being aware of Eve is $\frac{2}{5} \times \frac{1}{2} + \frac{1}{5} \times \frac{1}{2} \times \frac{1}{2} = \frac{1}{4}$. Obviously, the failure probability is  $1 - {\left( {\frac{3}{4}} \right)^m}$ when $m$ particle pairs are attacked.

\subsubsection{Measure-resend attack}

Eve intercepts sequence ${S_{dp(i)}}$ $\left( {{S_{r(i)}}} \right)$  and measure the particles in  ${Z_{dp}}$ $\left( {{Z_r}} \right)$ basis. Eve generates fake particles according to the measurement results and sends them to TP. The states of the particles in ${X_{dp}}$ $\left( {{X_r}} \right)$ basis will be changed by Eve’s actions in \textbf{Case 1 (Case 1*)}. If the original particle pair sent from TP is in ${Z_{dp}}$ $\left( {{Z_r}} \right)$ basis, then Eve will not be perceived. If the particle pair sent from TP is in ${X_{dp}}$$\left( {{X_r}} \right)$ basis, Eve will be detected with a probability of $\frac{1}{2}$. The probability of discovering Eve’s attack is $\frac{1}{5} \times \frac{1}{2} \times \frac{1}{2} = \frac{1}{{20}}$. Therefore, the total detection rate of the proposed protocol is $1 - {\left( {\frac{{19}}{{20}}} \right)^m}$.

\subsubsection{Entanglement attack}

Eve could entangle the auxiliary state $\left| \varepsilon  \right\rangle$ on the particles with unitary operation ${U_E}$ . Then Eve measures the auxiliary states to filch the private secrets of classical participants. The performance of resisting the entanglement attack of the first MSQPC protocol will be analyzed. The similar conclusion could also be obtained for the second MSQPC protocol. If Eve executes unitary operation ${U_E}$, the particles will evolve into

\begin{equation}
\begin{array}{c}
{U_E}\left| {{0_{dp}}} \right\rangle \left| \varepsilon  \right\rangle  = {\lambda _{00}}\left| {00} \right\rangle \left| {{E_{00}}} \right\rangle  + {\lambda _{01}}\left| {01} \right\rangle \left| {{E_{01}}} \right\rangle  + {\lambda _{10}}\left| {10} \right\rangle \left| {{E_{10}}} \right\rangle  + {\lambda _{11}}\left| {11} \right\rangle \left| {{E_{11}}} \right\rangle; \\
{U_E}\left| {{1_{dp}}} \right\rangle \left| \varepsilon  \right\rangle  = {\omega _{00}}\left| {00} \right\rangle \left| {E_{00}^{'}} \right\rangle  + {\omega _{01}}\left| {01} \right\rangle \left| {E_{01}^{'}} \right\rangle  + {\omega _{10}}\left| {10} \right\rangle \left| {E_{10}^{'}} \right\rangle  + {\omega _{11}}\left| {11} \right\rangle \left| {E_{11}^{'}} \right\rangle; \\
{U_E}\left| {{ + _{dp}}} \right\rangle \left| \varepsilon  \right\rangle  = \frac{1}{{\sqrt 2 }}\left( {{U_E}\left| {{0_{dp}}} \right\rangle \left| \varepsilon  \right\rangle  + {U_E}\left| {{1_{dp}}} \right\rangle \left| \varepsilon  \right\rangle } \right)\\
= \frac{1}{2}\left[ \begin{array}{c}
\left| {{\phi ^ + }} \right\rangle \left( {{\lambda _{00}}\left| {{E_{00}}} \right\rangle  + {\lambda _{11}}\left| {{E_{11}}} \right\rangle  + {\omega _{00}}\left| {E_{00}^{'}} \right\rangle  + {\omega _{01}}\left| {E_{01}^{'}} \right\rangle } \right)\\
+ \left| {{\phi ^ - }} \right\rangle \left( {{\lambda _{00}}\left| {{E_{00}}} \right\rangle  - {\lambda _{10}}\left| {{E_{10}}} \right\rangle  + {\omega _{00}}\left| {E_{00}^{'}} \right\rangle  - {\omega _{01}}\left| {E_{01}^{'}} \right\rangle } \right)\\
+ \left| {{\psi ^ + }} \right\rangle \left( {{\lambda _{01}}\left| {{E_{00}}} \right\rangle  + {\lambda _{11}}\left| {{E_{11}}} \right\rangle  + {\omega _{01}}\left| {E_{01}^{'}} \right\rangle  + {\omega _{10}}\left| {E_{10}^{'}} \right\rangle } \right)\\
+ \left| {{\psi ^ - }} \right\rangle \left( {{\lambda _{01}}\left| {{E_{00}}} \right\rangle  - {\lambda _{11}}\left| {{E_{11}}} \right\rangle  + {\omega _{01}}\left| {E_{01}^{'}} \right\rangle  - {\omega _{10}}\left| {E_{10}^{'}} \right\rangle } \right)
\end{array} \right];\\
{U_E}\left| {{ - _{dp}}} \right\rangle \left| \varepsilon  \right\rangle  = \frac{1}{{\sqrt 2 }}\left( {{U_E}\left| {{0_{dp}}} \right\rangle \left| \varepsilon  \right\rangle  - {U_E}\left| {{1_{dp}}} \right\rangle \left| \varepsilon  \right\rangle } \right)\\
= \frac{1}{2}\left[ \begin{array}{c}
\left| {{\phi ^ + }} \right\rangle \left( {{\lambda _{00}}\left| {{E_{00}}} \right\rangle  + {\lambda _{11}}\left| {{E_{11}}} \right\rangle  - {\omega _{00}}\left| {E_{00}^{'}} \right\rangle  - {\omega _{01}}\left| {E_{01}^{'}} \right\rangle } \right)\\
+ \left| {{\phi ^ - }} \right\rangle \left( {{\lambda _{00}}\left| {{E_{00}}} \right\rangle  - {\lambda _{10}}\left| {{E_{10}}} \right\rangle  - {\omega _{00}}\left| {E_{00}^{'}} \right\rangle  + {\omega _{01}}\left| {E_{01}^{'}} \right\rangle } \right)\\
+ \left| {{\psi ^ + }} \right\rangle \left( {{\lambda _{01}}\left| {{E_{00}}} \right\rangle  + {\lambda _{11}}\left| {{E_{11}}} \right\rangle  - {\omega _{01}}\left| {E_{01}^{'}} \right\rangle  - {\omega _{10}}\left| {E_{10}^{'}} \right\rangle } \right)\\
+ \left| {{\psi ^ - }} \right\rangle \left( {{\lambda _{01}}\left| {{E_{00}}} \right\rangle  - {\lambda _{11}}\left| {{E_{11}}} \right\rangle  + {\omega _{01}}\left| {E_{01}^{'}} \right\rangle  + {\omega _{10}}\left| {E_{10}^{'}} \right\rangle } \right)
\end{array} \right].\\

\end{array}
\end{equation}
where ${\left| {{\lambda _{00}}} \right|^2} + {\left| {{\lambda _{01}}} \right|^2} + {\left| {{\lambda _{10}}} \right|^2} + {\left| {{\lambda _{11}}} \right|^2} = {\left| {{\omega _{00}}} \right|^2} + {\left| {{\omega _{01}}} \right|^2} + {\left| {{\omega _{10}}} \right|^2} + {\left| {{\omega _{11}}} \right|^2} = 1$. If Eve attempts to pass the detection, ${\lambda _{00}} = {\lambda _{10}} = {\lambda _{11}} = {\omega _{00}} = {\omega _{01}} = {\omega _{11}} = 0$ and ${\lambda _{11}}\left| {{E_{11}}} \right\rangle  = {\omega _{01}}\left| {E_{01}^{'}} \right\rangle $ must be satisfied. However, Eve couldn’t distinguish the auxiliary states ${\lambda _{01}}\left| {{E_{01}}} \right\rangle $ and ${\omega _{10}}\left| {E_{10}^{'}} \right\rangle $. In other words, Eve couldn’t learn any valuable information by measuring the auxiliary states. Hence, Eve’s entanglement attack will fail.

\subsubsection{Trojan horse attack}

Eve could execute the Trojan horse attacks to filch the private information, since the particles are transmitted back and forth. Hence, all the legitimate participants should equip with wavelength quantum filters and photon number splitters.

\subsection{Inside attack}

Unlike outside attackers, internal participants seem to have strong eavesdropping capabilities \cite{26}. The attacks from semi-honest TP and dishonest participant will be analyzed.

\subsubsection{Semi-honest TP’s attack}

Semi-honest TP has more chance to obtain information than other attackers since he is both the sender and the receiver of particles. Suppose TP is dishonest and he attempts to obtain the privacy of ${P_i}$ , he will execute any attack but couldn’t cooperate with others. TP could extract the private information from ${r_{i,j}}$ $\left( {r_{i,j}^*} \right)$. Fortunately, TP couldn’t deduce any valuable information since $n$ participants share the common key $K$ $\left( {{K^*}} \right)$ in advance.

\subsubsection{Dishonest participants’ attack}

The roles of participants in the proposed protocols are equivalent. Assume that ${P_1}$  is dishonest and tries to acquire the other participants’ personal information. If ${P_1}$  executes attacks, he will be detected as an external stealer as explicated in \textbf {Section 3.1}. That is to say, our two protocols could withstand the dishonest participant’s attack.

\section{Simulation}\label{sec.4}

Based on the IBM Quantum Experience \cite{27}, the operations CTRL and SIFT involved in the proposed protocols are simulated.

\subsection{Simulation analysis on the first MSQPC protocol}

With the Bell measurement in ${X_{dp}}$basis, the relative measurements results are

\begin{equation}
\begin{array}{c}
{\left| {{ + _{dp}}} \right\rangle _{12}} = \frac{1}{{\sqrt 2 }}{\left( {\left| {01} \right\rangle  + \left| {10} \right\rangle } \right)_{12}}\overset{{{\rm{CNO}}{{\rm{T}}_{12}}}}{\rightarrow} \frac{1}{{\sqrt 2 }}{\left( {\left| {01} \right\rangle  + \left| {11} \right\rangle } \right)_{12}}\overset{{{H_1}}}{\rightarrow} {\left| {01} \right\rangle _{12}},\\
{\left| {{ - _{dp}}} \right\rangle _{12}} = \frac{1}{{\sqrt 2 }}{\left( {\left| {01} \right\rangle  - \left| {10} \right\rangle } \right)_{12}}\overset{{{\rm{CNO}}{{\rm{T}}_{12}}}}{\rightarrow} \frac{1}{{\sqrt 2 }}{\left( {\left| {01} \right\rangle  - \left| {11} \right\rangle } \right)_{12}}\overset{{{H_1}}}{\rightarrow} {\left| {11} \right\rangle _{12}}.
\end{array}
\end{equation}
In \textbf{Step 1}, TP should prepare the DF logical states and send them to ${P_i}$ , the DF logical states can be prepared as

\begin{equation}
\begin{array}{c}
{\left| {00} \right\rangle _{12}}\overset{{{I_1} \otimes {X_2}}}{\rightarrow} {\left| {01} \right\rangle _{12}} = \left| {{0_{dp}}} \right\rangle; \\
{\left| {00} \right\rangle _{12}}\overset{{{X_1} \otimes {I_2}}}{\rightarrow} {\left| {10} \right\rangle _{12}} = \left| {{1_{dp}}} \right\rangle; \\
{\left| {00} \right\rangle _{12}}\overset{{{H_1} \otimes {X_2}}}{\rightarrow} {\left| { + 1} \right\rangle _{12}} \overset{{\rm{CNO}}{{\rm{T}}_{12}}}{\rightarrow}  \frac{1}{{\sqrt 2 }}{\left( {\left| {01} \right\rangle  + \left| {10} \right\rangle } \right)_{12}} = \left| {{ + _{dp}}} \right\rangle; \\
{\left| {00} \right\rangle _{12}}\overset{{{X_1} \otimes {X_2}}}{\rightarrow} {\left| {11} \right\rangle _{12}}\overset{{{H_1} \otimes {I_2}}}{\rightarrow} {\left| { - 1} \right\rangle _{12}} \overset{{\rm{CNO}}{{\rm{T}}_{12}}}{\rightarrow}  \frac{1}{{\sqrt 2 }}{\left( {\left| {01} \right\rangle  - \left| {10} \right\rangle } \right)_{12}} = \left| {{ - _{dp}}} \right\rangle .
\end{array}
\end{equation}

\subsubsection{CTRL operation}

Assume that the original DF logical qubits distributed by TP are in $\left| {{ - _{dp}}} \right\rangle$. TP will measure the particles in ${X_{dp}}$ basis in \textbf{Case 1} if ${P_i}$ executes the CTRL operation on particles in \textbf{Step 2}. There are two cases in the CTRL operation. Besides, the collective-dephasing noise could be simulated by ${\rm{RZ}}\left( {\frac{\pi }{5}} \right)$.

\begin{figure}
	\centering
	\includegraphics[width=1\linewidth]{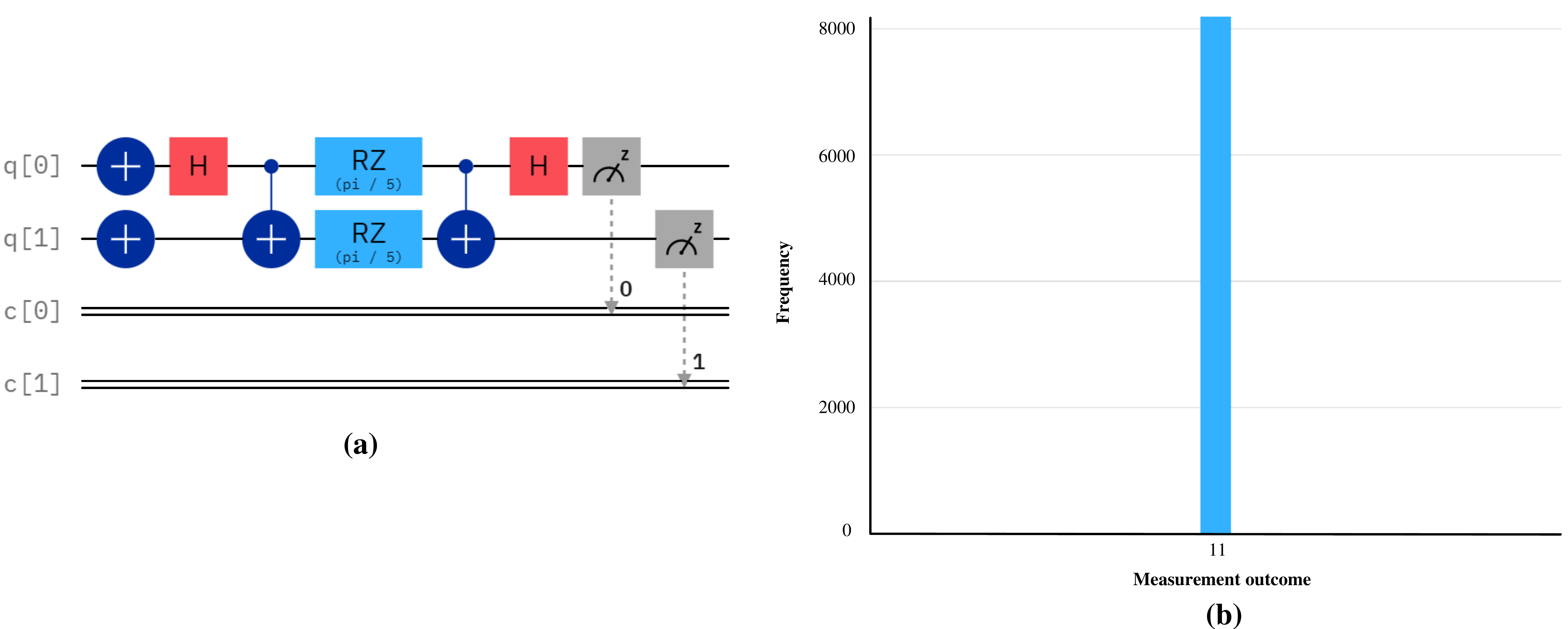}
	\caption{\textbf{(a)} Quantum circuit of CTRL operation, \textbf{(b)} the measurement results in a secure quantum channel with the first MSQPC protocol}
	\label{Fig1.}
\end{figure}

(1) As for the secure quantum channel, the quantum circuit and the corresponding measurement results are described in Figure \ref{Fig1.}. TP’s measurement outcome is $\left| {11} \right\rangle $, which signifies that the result of ${X_{dp}}$ basis measurement is $\left| {{ - _{dp}}} \right\rangle $.

\begin{equation}
\begin{array}{c}
\left| {{ - _{dp}}} \right\rangle  = \frac{1}{{\sqrt 2 }}{\left( {\left| {01} \right\rangle  - \left| {10} \right\rangle } \right)_{12}}\overset{{{\rm{RZ}}\left( {\frac{\pi }{5}} \right)}}{\rightarrow} \frac{{{e^{i\frac{\pi }{5}}}}}{{\sqrt 2 }}{\left( {\left| {01} \right\rangle  - \left| {10} \right\rangle } \right)_{12}}\\
\overset{{{\rm{CNO}}{{\rm{T}}_{12}}}}{\rightarrow} \frac{{{e^{i\frac{\pi }{5}}}}}{{\sqrt 2 }}{\left( {\left| {01} \right\rangle  - \left| {11} \right\rangle } \right)_{12}}\overset{{{H_1}}}{\rightarrow} {e^{i\frac{\pi }{5}}}{\left| {11} \right\rangle _{12}}.
\end{array}
\end{equation}

\begin{figure}
	\centering
	\includegraphics[width=1\linewidth]{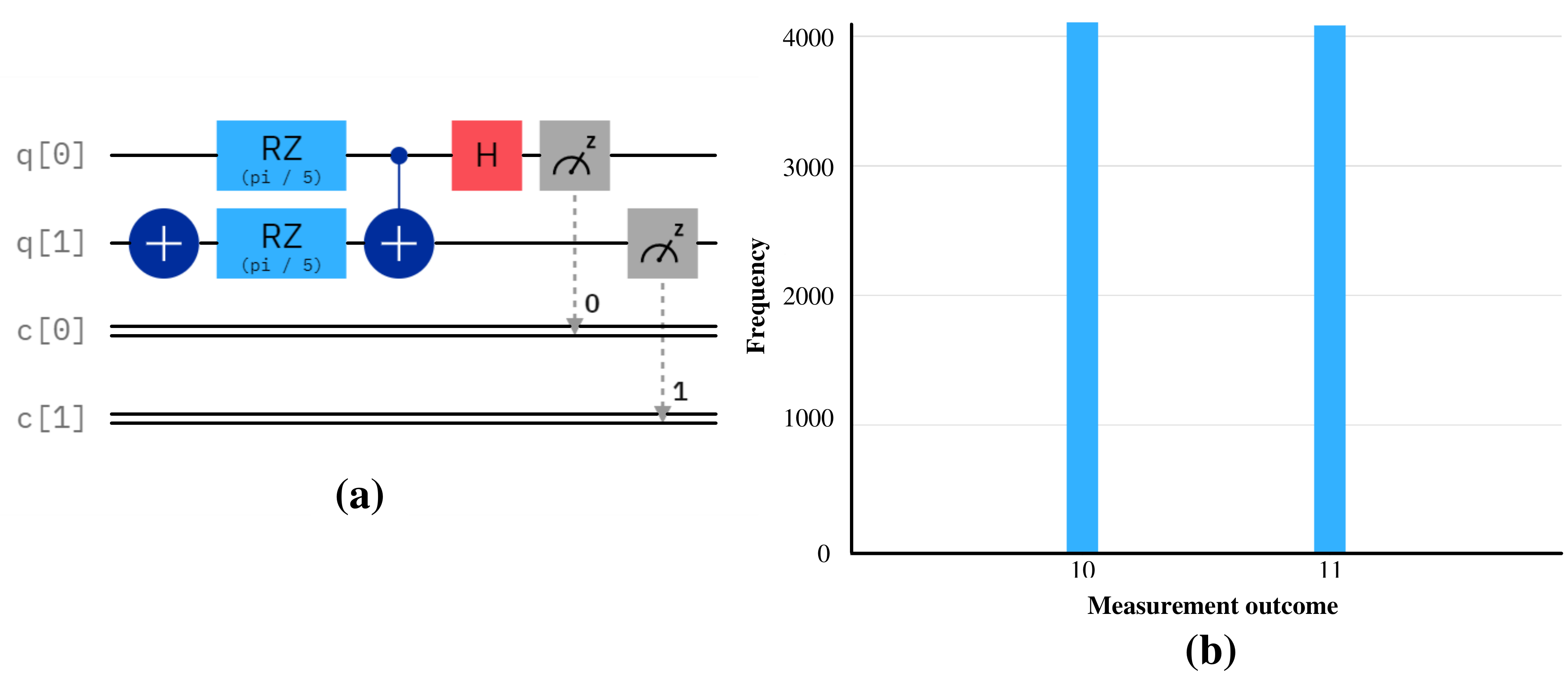}
	\caption{\textbf{(a)} Quantum circuit of CTRL operation, \textbf{(b)} the measurement results in a insecure quantum channel with the first MSQPC protocol}
	\label{Fig2.}
\end{figure}

(2) As for the insecure quantum channel, Eve may launch intercept-resend attack or measure-resend attack. Supposed that the fake particles distributed by Eve are in $\left| {{0_{dp}}} \right\rangle$, the simulation of Eve’s attacks is indicated in Figure \ref{Fig2.}. TP’s measurement outcome is random in $\left| {01} \right\rangle $ or $\left| {11} \right\rangle $. As long as Eve executes attack actions, TP has the chance to detect Eve.

\begin{equation}
\begin{array}{c}
\left| {{0_{dp}}} \right\rangle  = {\left| {01} \right\rangle _{12}}\overset{{{\rm{RZ}}\left( {\frac{\pi }{5}} \right)}}{\rightarrow} {e^{i\frac{\pi }{5}}}{\left| {01} \right\rangle _{12}}\\
\overset{{{\rm{CNO}}{{\rm{T}}_{12}}}}{\rightarrow} {e^{i\frac{\pi }{5}}}{\left| {01} \right\rangle _{12}}\overset{{{H_1}}}{\rightarrow} \frac{{{e^{i\frac{\pi }{5}}}}}{{\sqrt 2 }}{\left( {\left| {01} \right\rangle  + \left| {11} \right\rangle } \right)_{12}}.
\end{array}
\end{equation}

\subsubsection{SIFT operation}
\begin{figure}
	\centering
	\includegraphics[width=1\linewidth]{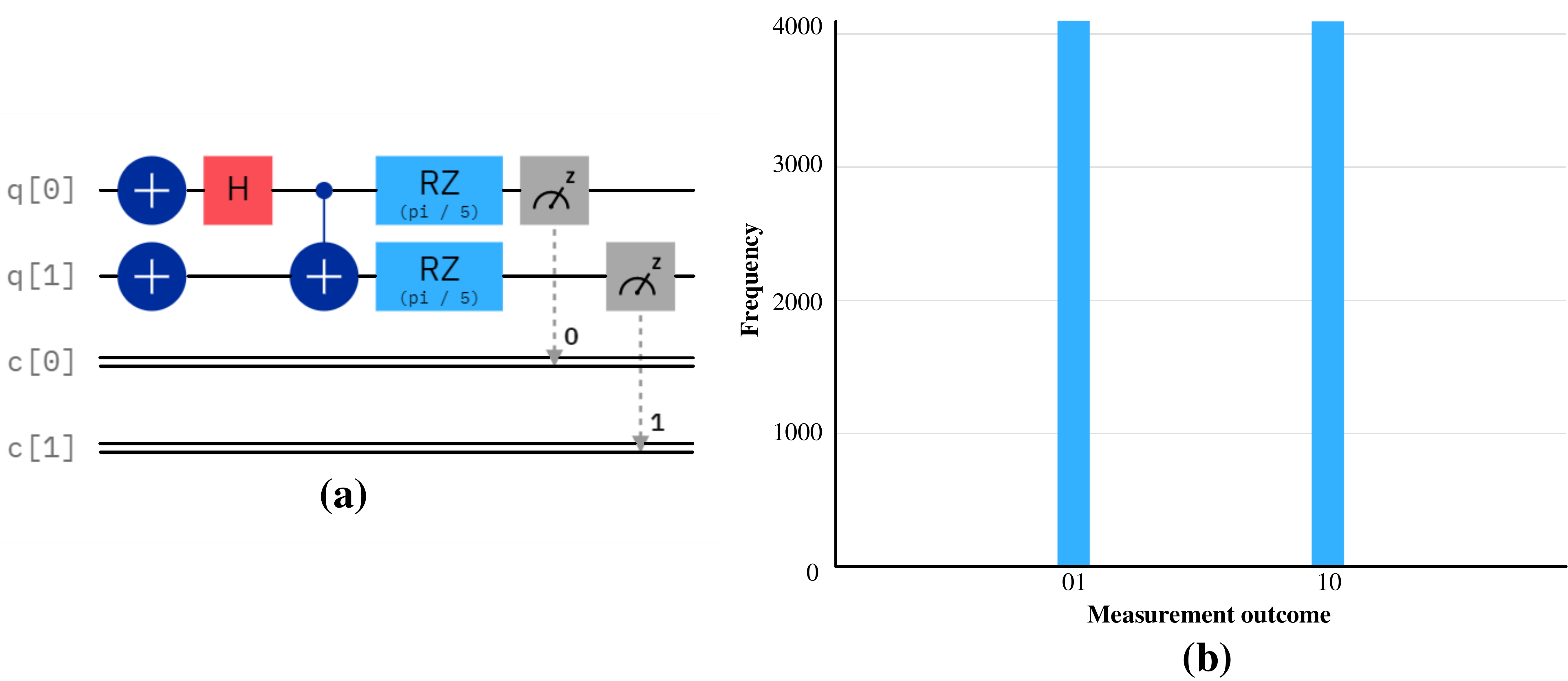}
	\caption{\textbf{(a)} Quantum circuit of SIFT operation, \textbf{(b)} the measurement results of the first MSQPC protocol}
	\label{Fig3.}
\end{figure}

If ${P_i}$  performs SIFT operation on particles in \textbf{Step 2}, the quantum circuit and the measurement results are shown in Figure \ref{Fig3.}. The finial measurement outcome of ${P_i}$  is random in $\left| {01} \right\rangle$ or $\left| {10} \right\rangle $.

\begin{equation}
\left| {{ - _{dp}}} \right\rangle  = \frac{1}{{\sqrt 2 }}{\left( {\left| {01} \right\rangle  - \left| {10} \right\rangle } \right)_{12}}\overset{{{\rm{RZ}}\left( {\frac{\pi }{5}} \right)}}{\rightarrow} \frac{{{e^{i\frac{\pi }{5}}}}}{{\sqrt 2 }}{\left( {\left| {01} \right\rangle  - \left| {10} \right\rangle } \right)_{12}}.
\end{equation}

\subsection{Simulation analysis on the second MSQPC protocol}

${Z_r}$ basis measurements could be achieved with $Z \otimes Z$  basis measurements since the parities of $\left| {{0_r}} \right\rangle $ and $\left| {{1_r}} \right\rangle $ are different and could be distinguished with two single-photon measurements easily. Besides, ${X_r}$ basis measurements could be realized by $H$ operation and $Z \otimes Z$ basis measurements. The results of ${X_r}$ basis measurements are

\begin{equation}
\begin{array}{c}
\left| {{ + _r}} \right\rangle  = \frac{1}{2}{\left( {\left| {00} \right\rangle  + \left| {11} \right\rangle  + \left| {01} \right\rangle  - \left| {10} \right\rangle } \right)_{12}}\overset{{{H_2}}}{\rightarrow} \frac{1}{{\sqrt 2 }}{(\left| {00} \right\rangle  + \left| {11} \right\rangle )_{12}};\\
\left| {{ - _r}} \right\rangle  = \frac{1}{2}{\left( {\left| {00} \right\rangle  + \left| {11} \right\rangle  - \left| {01} \right\rangle  + \left| {10} \right\rangle } \right)_{12}}\overset{{{H_2}}}{\rightarrow} \frac{1}{{\sqrt 2 }}{(\left| {01} \right\rangle  + \left| {10} \right\rangle )_{12}}.
\end{array}
\end{equation}
In \textbf{Step 1*}, TP prepares the DF logical states as follows,

\begin{equation}
\begin{array}{c}
{\left| {00} \right\rangle _{12}}\overset{{{H_1} \otimes {I_2}}}{\rightarrow} {\left| { + 0} \right\rangle _{12}}\overset{{{\rm{CNO}}{{\rm{T}}_{12}}}}{\rightarrow} \frac{1}{{\sqrt 2 }}{\left( {\left| {00} \right\rangle  + \left| {11} \right\rangle } \right)_{12}} = \left| {{0_r}} \right\rangle; \\
{\left| {00} \right\rangle _{12}}\overset{{{X_1} \otimes {X_2}}}{\rightarrow} {\left| {11} \right\rangle _{12}}\overset{{{H_1} \otimes {I_2}}}{\rightarrow} {\left| { - 1} \right\rangle _{12}}\overset{{{\rm{CNO}}{{\rm{T}}_{12}}}}{\rightarrow} \frac{1}{{\sqrt 2 }}{\left( {\left| {01} \right\rangle  - \left| {10} \right\rangle } \right)_{12}} = \left| {{1_r}} \right\rangle ;\\
{\left| {00} \right\rangle _{12}}\overset{{{H_1} \otimes {X_2}}}{\rightarrow} {\left| { + 1} \right\rangle _{12}}\overset{{{\rm{CNO}}{{\rm{T}}_{12}}}}{\rightarrow} \frac{1}{{\sqrt 2 }}{\left( {\left| {01} \right\rangle  + \left| {10} \right\rangle } \right)_{12}}\overset{{{H_1}}}{\rightarrow} \\
\frac{1}{{\sqrt 2 }}{\left( {\left| {00} \right\rangle  + \left| {11} \right\rangle  + \left| {01} \right\rangle  - \left| {10} \right\rangle } \right)_{12}} = \left| {{ + _r}} \right\rangle; \\
{\left| {00} \right\rangle _{12}}{\left| {10} \right\rangle _{12}}{\left| { - 0} \right\rangle _{12}}\frac{1}{{\sqrt 2 }}{\left( {\left| {00} \right\rangle  - \left| {11} \right\rangle } \right)_{12}}\\
\frac{1}{{\sqrt 2 }}{\left( {\left| {00} \right\rangle  + \left| {11} \right\rangle  - \left| {01} \right\rangle  + \left| {10} \right\rangle } \right)_{12}} = \left| {{ - _r}} \right\rangle.
\end{array}
\end{equation}

\subsubsection{CTRL operation}

Suppose the original logical qubits sent by TP are in $\left| {{ - _r}} \right\rangle $ . If ${P_i}$  operates the CTRL, TP will measure the logical qubits in ${X_r}$ basis in \textbf{Case 1*}. There are two cases in the CTRL operation. The collective-rotation noise could be simulated by ${\rm{RY}}\left( {\frac{\pi }{5}} \right)$.

\begin{figure}
	\centering
	\includegraphics[width=1\linewidth]{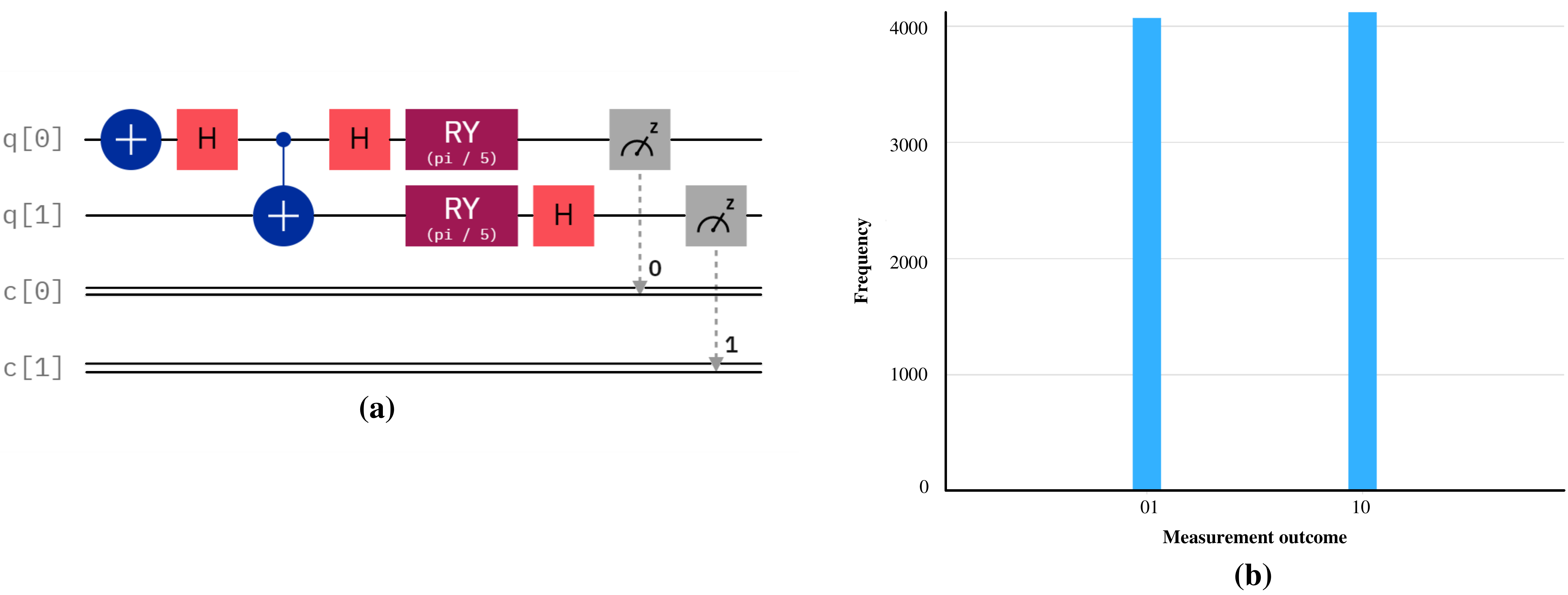}
	\caption{\textbf{(a)} Quantum circuit of CTRL operation, \textbf{(b)} the measurement results in a secure quantum channel of the second MSQPC protocol}
	\label{Fig4.}
\end{figure}

(1) As for the secure quantum channel, the quantum circuit and the measurement results are shown in Figure \ref{Fig4.}. TP’s measurement outcome is random in $\left| {01} \right\rangle$ or $\left| {10} \right\rangle$, which implies the result of ${X_r}$ basis measurement is $\left| {{ - _r}} \right\rangle$.

\begin{equation}
\begin{array}{c}
\left| {{ - _r}} \right\rangle  = \frac{1}{2}{\left( {\left| {00} \right\rangle  + \left| {11} \right\rangle  - \left| {01} \right\rangle  + \left| {10} \right\rangle } \right)_{12}}\\
\overset{{{\rm{RY}}\left( {\frac{\pi }{5}} \right)}}{\rightarrow} \frac{1}{2}{\left( {\left| {00} \right\rangle  + \left| {11} \right\rangle  - \left| {01} \right\rangle  + \left| {10} \right\rangle } \right)_{12}}\overset{{{H_2}}}{\rightarrow}\frac{1}{2}{\left( {\left| {01} \right\rangle  + \left| {10} \right\rangle } \right)_{12}}.
\end{array}
\end{equation}

\begin{figure}
	\centering
	\includegraphics[width=1\linewidth]{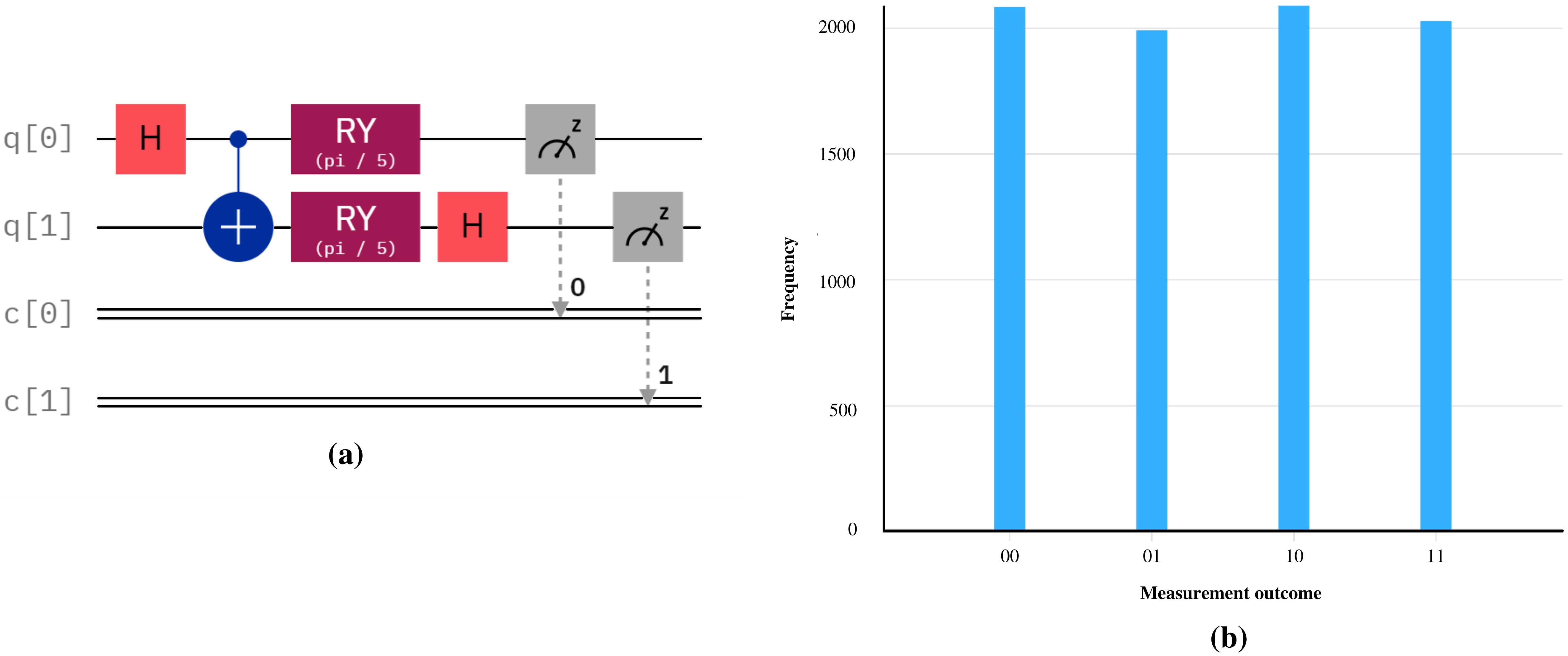}
	\caption{\textbf{(a)} Quantum circuit of CTRL operation, \textbf{(b)} the measurement results in an insecure quantum channel of the second MSQPC protocol}
	\label{Fig5.}
\end{figure}

(2) When the quantum channel is insecure, assume that the fake particles sent by Eve are in $\left| {{0_r}} \right\rangle$, Figure \ref{Fig5.} describes the simulation of Eve’s attacks. TP’s measurement outcome is random in $\left| {00} \right\rangle$, $\left| {01} \right\rangle$, $\left| {10} \right\rangle$ or $\left| {11} \right\rangle$. If TP’s measurement result is $\left| {00} \right\rangle$ or $\left| {11} \right\rangle$, he could know that the quantum channel is insecure.

\begin{equation}
\begin{array}{c}
\left| {{0_r}} \right\rangle  = \frac{1}{{\sqrt 2 }}{\left( {\left| {00} \right\rangle  + \left| {11} \right\rangle } \right)_{12}}\overset{{{\rm{RY}}\left( {\frac{\pi }{5}} \right)}}{\rightarrow} \frac{1}{{\sqrt 2 }}{\left( {\left| {00} \right\rangle  + \left| {11} \right\rangle } \right)_{12}}\\
\overset{{{H_2}}}{\rightarrow}\frac{1}{{\sqrt 2 }}{\left( {\left| {00} \right\rangle  + \left| {01} \right\rangle  + \left| {10} \right\rangle  - \left| {11} \right\rangle } \right)_{12}}.
\end{array}.
\end{equation}

\subsubsection{SIFT operation}

\begin{figure}
	\centering
	\includegraphics[width=1\linewidth]{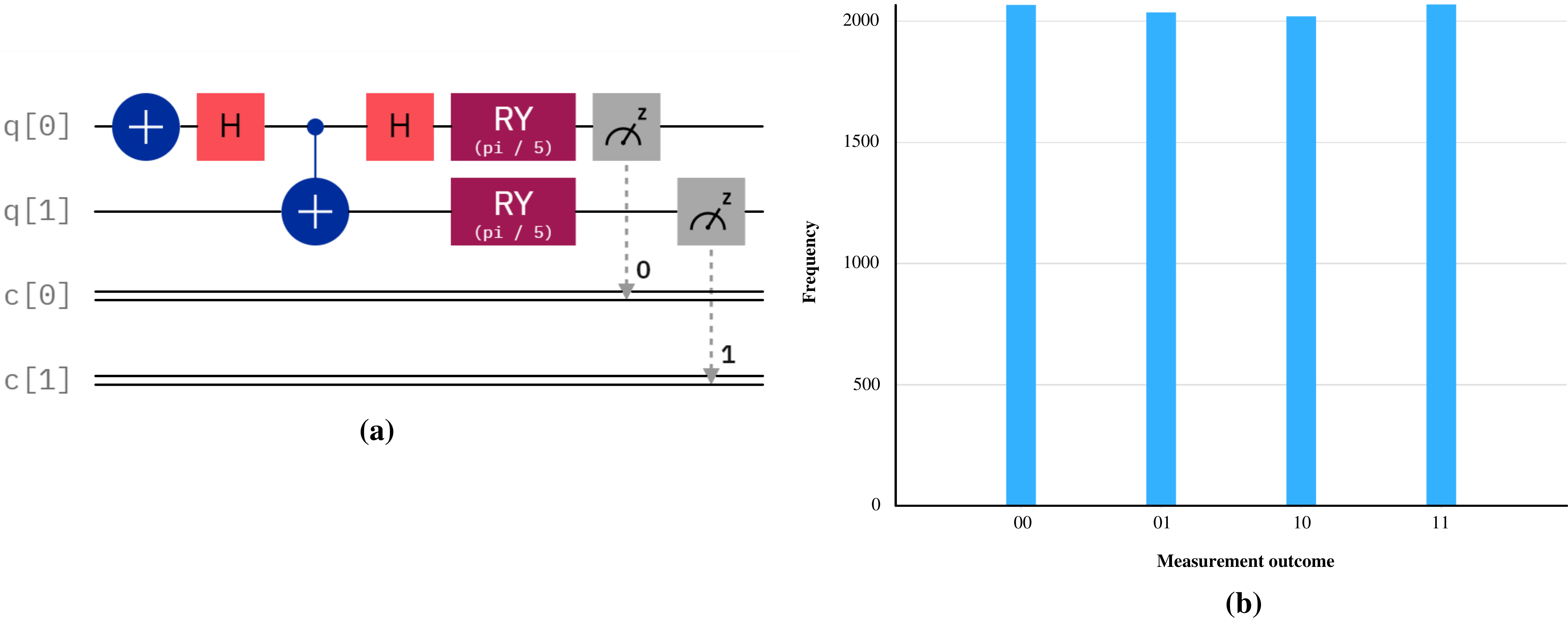}
	\caption{\textbf{(a)} Quantum circuit of SIFT operation, \textbf{(b)} the measurement results in the secure quantum channel of the second MSQPC protocol}
	\label{Fig6.}
\end{figure}

If ${P_i}$  performs SIFT operation, the quantum circuit and the measurement results are shown in Figure \ref{Fig6.}. The measurement outcome of ${P_i}$  is random in one of the states $\left| {00} \right\rangle$, $\left| {01} \right\rangle$, $\left| {10} \right\rangle$ or $\left| {11} \right\rangle$, which is consistent with Equation \eqref{Eq1.}.

\begin{equation}
\left| {{ - _r}} \right\rangle  = \frac{1}{2}{\left( {\left| {00} \right\rangle  + \left| {11} \right\rangle  + \left| {01} \right\rangle  - \left| {10} \right\rangle } \right)_{12}}\overset{{{\rm{RY}}\left( {\frac{\pi }{5}} \right)}}{\rightarrow}\frac{1}{2}{\left( {\left| {00} \right\rangle  + \left| {11} \right\rangle  + \left| {01} \right\rangle  - \left| {10} \right\rangle } \right)_{12}}
\label{Eq1.}
\end{equation}

\section{Comparison}\label{sec.5}

 The detailed comparison between the two proposed protocols and some baseline protocols is collected in Table \ref{Table 3}. Similar with the protocols in \cite{21,22,23}, our protocols could resist the collective noises. Besides, the participants in our protocols are semi-quantum and not necessary to own full quantum capability. Like the protocols in \cite{20,23}, our protocol could compute the private information of $n$  the participants, while the schemes in \cite{21,23} only could compare two participants’ privacies. Furthermore, all the classical participants in our protocols should share a common key.
 
 The qubit efficiency of quantum protocols could be represented as $\xi  = c/t$ \cite{28}, where $c$ and $t$ represent the number of compared bits and the total number of qubits prepared in the protocols, respectively. In the proposed protocols, the length of secrets is $l$, hence $c = nl$. TP should prepare $5nl$ DF logical states, equivalently, TP should generate $10nl$ qubits. Classical participants will execute SIFT operation with a probability of $\frac{1}{2}$, and they will prepare $5nl$ qubits. Hence, the qubit efficiency of our protocols is $\xi  = \frac{{nl}}{{10nl + 5nl}} = \frac{1}{{15}}$. Semi-quantum protocols require more quantum resources since the classical participants only have restricted quantum capabilities. Besides, it is necessary to consume more quantum resources to resist noises. Though the proposed protocols have no advantage in terms of quantum efficiency, they could work in the semi-quantum mode and could resist noises effectively. 

\begin{table}[!ht]
	\addtolength{\leftskip}{-4cm}
		\caption{Comparison}
		\begin{tabular}{  m{2cm}<{\centering} m{2.5cm}<{\centering}  m{3cm}<{\centering}  m{3cm}<{\centering} m{2.5cm}<{\centering} m{3cm}<{\centering} m{1.8cm}<{\centering} } 	
			\hline
			& Category of QPC & Number of participants &  Quantum resource & Need for pre-shared key & Channel type & Quantum efficiency \\ 
			\hline
		  \cite{20} & SQPC & $n(n \ge 3)$ &  Bell state & Yes & Idea channel & $\frac{1}{5}$ \\ 
          \cite{21} & Noise-resisting QPC & 2 &  GHZ state & No & Collective noise channel & $\frac{2}{15}$ \\ 
          \cite{22} & Noise-resisting QPC & $n(n \ge 3)$ &  Three-qubit entangled state & Yes & Collective noise channel & $\frac{1}{3}$ \\ 
          \cite{23} & Noise-resisting SQPC & 2 & DF state & No & Collective noise channel & $\frac{1}{48}$ \\ 
          Our protocols & Noise-resisting SQPC & $n(n \ge 3)$ &  DF state & Yes & Collective noise channel & $\frac{1}{15}$ \\ 
			\hline	
			\label{Table 3}
		\end{tabular}
\end{table}

\section{Conclusion}\label{sec.6}

To counteract collective noises, two robust MSQPC protocols are investigated with the DF states. Our two protocols have good performance in resisting collective-dephasing noise and collective-rotation noise, respectively. Assisting by a semi-honest TP with full quantum ability, multiple classical participants could compare their privacies. Besides, the proposed MSQPC protocols perform well in resisting various external attacks and internal attacks. Furthermore, the two proposed protocols are feasible under the related quantum facilities and technology, since the operations involving in our protocols could be implemented on the IBM Quantum Experience.

\section*{Acknowledgments}
This work is supported by the National Natural Science Foundation of China (Grant Nos. 62161025 and 61871205), and the Top Double 1000 Talent Programme of Jiangxi Province (Grant No. JXSQ2019201055).

\end{document}